\documentstyle[12pt]{article}
\begin{document}
\begin{center}
{\Large  Planck scale still safe from stellar images}\\
\bigskip
{\bf D.H. Coule}\\
\bigskip
Institute of Cosmology and Gravitation,\\ University of
Portsmouth, Mercantile House, Hampshire Terrace, Portsmouth PO1
2EG.\\
\bigskip

\begin{abstract}
The recent paper of Lieu and Hillman [1]  that a possible,
(birefringence like) phase difference ambiguity coming from Planck
effects would alter stellar images of distant sources  is
questioned. Instead for {\em division of wavefront} interference
and diffraction phenomena, initial (lateral) coherence is
developed simply by propagation of rays ( cf.  van Cittert-Zernike
theorem). This case is strongly immune to quantum gravity
influences that could tend to reduce phase coherence. The phase
ambiguity, if actually present, could reduce  any underlying
polarization of the light rays.

However,  we argue that, as expected since any  inherent  quantum
discreetness of space should become  increasingly negligible over
larger distances,  such a phase ambiguity is rapidly cancelled  if
a  more realistic  constantly fluctuating quantum  ``buffeting''
occurs.
\\


\end{abstract}
\end{center}
\newpage
{\bf 1.0 Introduction }\\
 In a recent paper Lieu and Hillman [1] have
suggested that the presence of interference effects in stellar
interferometers is contrary to what should be expected if time
fails to be defined below the Planck time $t_{pl} \sim 10^{-44}
sec$. These results, if true, would have profound consequences for
quantum gravity research as emphasized in press accounts of this
work, including {\em The Times} of  London  (8th Feb, 2003).

 In their analysis they consider the phase change of waves that occur on
 propagating through space from  distant objects.
  Because of quantum uncertainties in
 time, this phase  can become so uncertain ( to $\sim 2\pi$ ) after
 travelling vast distances $\sim$ {\em  Mpc}  in space
  that clear interference patterns should    no longer be possible. However we shall argue that  the authors
 have neglected an opposing property of waves they they simply gain coherence by propagating, which counteracts against
 any loss of phase caused by quantum gravitational fluctuations.

  However, rather than affecting interference or
  diffraction patterns significantly
  this phase could   alter  the possible polarization state of any observed
  light. Although this alone would be extremely important we
   find little justification  for such a random phase term being present due to
   quantum gravity discreteness alone.\\
 {\bf 2.0 Michelson Interferometer}\\
To clarify some of these issues we first consider, at face value,
the phase ambiguity outlined in ref.[1]. This   would actually
 be relevant for a Michelson interferometer -see eg[2-4] . Recall there that a light
 source is split by means of a partially reflecting mirror. These
 beams are then reflected back and depending on the relevant phase
 difference between the beams, interference fringes are seen. In the language of ref.[2] this is an example of
 {\em division of amplitude} interference phenomena . Such an arrangement was
 originally used to rule out motion through the ``ether''. Presumably if
 the arms of the interferometer were of $\sim${\em Mpc} scale then an
 uncertainty in phase would prevent longitudinal coherence and such fringes
would no longer be present. Of course this arrangement is
impossible to make, and has merely theoretical interest  in
placing an upper bound on  allowed coherence. However,  future
generations of space born gravitational wave ( of Michelson type)
interferometers, with ultra high frequency lasers together with
the possibility of a low energy quantum gravity scale $\sim Tev$
means this effect might possibly be amenable to investigation -
see ref.[5] for more detailed discussion of this possibility.
\\
     {\bf 3.0 Stellar interferometry}\\
The later developed  Michelson stellar interferometer or its
electronic Hanbury-Brown and Twiss  versions works on a different
principle [2,3]. It is an example of a {\em division of wavefront}
interference effect [2]. The lateral coherence between the
telescope's two apertures is the quantity now measured. The
diameter of the circular coherent area is given by $\sim 0.16
\lambda/\alpha$, where $\lambda$ is the wavelength and $\alpha$
the angle subtended at the aperture [2,3].  For a finite size star
this coherence only extends a finite distance which can be
measured by moving the mirrors/ receiving dishes until the
presence of interference fringes ( or their electronic analogues)
are no longer present. For the sun this distance is only $\sim
0.019mm$ while for the star Betelgeuse it is $\sim 2.5 m$ [2,3].
From this can be deduced the diameter of the source, typically
stars or bright galaxies. For a point source the interference
pattern is always present regardless of how far apart the mirrors
are placed. The calculation in this case is then the same as that
for Fraunhofer diffraction through two circular holes [2]. The
underlying phase is not important since the sources can be assumed
to be incoherent and any temporal or longitudinal coherence is
never required .

Now the only way this argument could be prevented is if the
initial transverse or spatial coherence is never present across
the two collecting mirrors. This is easily seen to be unreasonable
by means of the van Cittert-Zernike theorem which roughly is  the
principle that an electromagnetic wave becomes spatially coherent
simply by the process of propagation [2-4]. Technically the
calculation now  depends of differences between path lengths
between points on an incoherent source and at a far distance
receiving surface. Without providing the messy details the
required sum over path differences is rather immune to any
underlying fuzzy Planck scale cf. section 10.4.2 in ref.[2]. Any
underlying phase differences cancel out in the calculation since
we sum over a large number of paths. As mentioned, the sun,
although an incoherent source, develops spatial coherence albeit
only over a small scale by the time is reaches earth. Young's
fringes can still be produced provided the two slits are very
close within the region of spatial coherence. The fact that this
coherence develops at all means that any quantum effect causing
decoherence is subdominant  and counteracted. Since any possible
path through space can be split up into a series of shorter ones
cf. Huygen's principle [2,3] , the classical effect necessarily
dominates.

 If quantum gravity had dominated it would be analogous
to placing a finely ground glass plate in front of the slits in
Young's experiment using a coherent  beam-broadened laser source
[4,6] . This introduces  an irregular phase variation producing a
fragmented interference pattern . If the glass plate is then
slowly moved, to produce an ensemble average,  the usual Airy disc
interference pattern is made fuzzy due to a reduction in the {\em
mutual coherence function} [2-4,6]. However, on propagation
through space the classical effect must dominate if coherence from
the sun or
 nearby stars is actually produced - corresponding to an overall
  growth in the {\em mutual coherence function}. \\
 {\bf 4.0 Optical Birefringence of the vacuum }\\
We now wish to consider the work of ref.[1] more directly but
first we wish to review a number of issues  that seem relevant
before proceeding, particularly optical birefringence - see
eg.[2].

 Recall that
some crystals are birefringent which allows  the speed of
propagation to takes two values. This produces two orthogonal
beams: ordinary and extraordinary. A well known consequence  is
the double image seen when viewing an object through a slab of
Iceland Spar ($CaCO_{3}$). The phase difference that gradually
evolves between the two beams can alter any initial polarization .
This is the principle of the quarter wave or half wave plate that
is used to alter the polarization of say an initially vertically
polarized beam e.g.[2]. There are a number of related phenomena
like the Kerr and Pockels effects where applied electric fields
actually alter the indices of refraction [2]. One can also set up
a {\em division of amplitude} interference effect between the
ordinary and extraordinary rays. This is the phenomena of {\em
interference figures} as seen in a polarizing microscope [2].
Because the two beams are orthogonal one cannot directly set up
the interference effect . Rather one uses an {\em analyzer}
involving a Nicol Prism to finally obtain the interference figure
[2] . These and other general forms of optical activity generally
just  involve polarization effects, as first passing a laser
through some crystal to rotate its polarization has no effect on
any subsequently Airy disc diffraction pattern.

Studies in Lorentz violations in Electrodynamics have  considered
the possibility that the vacuum itself is birefringent so that
polarization changes or pulse dispersion might occur as waves pass
through space [7]. If such phenomena exist then initially
polarized light will gradually change as a phase shift develops
between the ordinary and extraordinary rays. In theory it might,
like in crystals,  be possible to create a interference figure
between the two rays although the type of analyzer now required is
unclear. Now the sort of analysis done by Lieu and Hillman, would
suggest that quantum gravity would make this phase difference
uncertain for large propagation lengths  so that polarization is
eventually lost and any future interference figure impossible to
produce. One might have argued that the still present polarization
in the Microwave background radiation is at odds from these
expected Planck scale degradation effects.

 However there are a number of
assumptions that can be questioned before we could reach such a
conclusion even if vacuum birefringence is present. Assume that
the two modes, travelling in the $z$  direction, have differing
phase velocities $v_p$ so that the relative phase $\Delta \phi$
between the x direction and y direction of the Electric field
changes by
\begin{equation}
\Delta \phi =2\pi \Delta v_p L/\lambda
\end{equation}
where $L$ is the distance travelled and $\lambda$ the wavelength.
So far this is deterministic and would simply alter any initial
polarization present. What we really require is for $\Delta \phi$
to become random which would signify that polarization is becoming
arbitrary after passing large distances.  Can the quantum gravity
discreteness produce this required randomness ?  It is unclear why
the phase velocity in say the x component should always receive
positive fluctuations $(+++++....)$ while those in the other y
component only negative fluctuations $(-----....)$ . One should
rather expect each ray to receive a random sequence e.g.
$(+-++-....)$. \footnote{The actual size of the fluctuation
determined by the parameters $a$ and $\alpha$ in refs.[1,8 ]is not
crucial for the following argument. } It isn't clear how often one
should consider a measurement to be made if at all. If we consider
N such measurements then the random sequence will typically take
the normalized  value $\pm N^{-1/2}$. Note that unlike a random
walk the velocity cannot keep wandering  further from its standard
value but is instead anchored, to within the confines of the
uncertainty principle, around $c$. In the limit $N\rightarrow
\infty$ the positive and negative fluctuation components cancel
out and no random phase component would be present. However, in
this notation ref. [1] have assumed $N=1$ so that the rays either
have a constantly smaller or greater than $c$ velocity throughout
their journey. Although this is one possible description of
presently unknown Planck scale physics it does not seem plausible.
In ref.[8] they have pointed out that if instead one takes a more
reasonable  $N=L/\lambda \sim 10^{30}$ any corresponding effect is
$\sim 10^{15}$ times smaller. But one could even envision taking
the Planck length to be the relevant scale of quantum
``buffeting'' so that the effect would now be around $\sim
10^{30}$  times smaller. Note that unless $N=1$ increasing the
distance $L$ reduces the cumulative effects of quantum
discreteness as more sampling is done. Especially since any actual
image is made up from large numbers of photons.  Only in the
birefringent limit $N=1$ does the fluctuation (phase difference)
grow with distance as in eq.(1). We would contend that a
reasonable interpretation of the effects of quantum foam is that a
large number of small buffetings occur during the passage through
space. Provided that this number $N>>1$ the effects of random
fluctuations  are rapidly cancelled out and don't accumulate in a
linear way with distance.

 If one is prepared to consider that the two
components of the electric field should be treated independently
during their travels  then it wouldn't be necessary to require
that the vacuum itself be birefringent. However since the two
electric field components are commuting variables this is
difficult to envision and  again any effect is rapidly reduced for
$N>>1$.

 {\bf 7.0  Conclusion}\\
 In conclusion the authors of ref.[1] have neglected an opposing
 property of propagating waves that tends to counteract any
phase  decoherence caused by underlying quantum gravitational
effects.

 Evading this principle with quantum gravity  is difficult
since it now depends on summing over all path length differences
between emitter and receiver, not a simple accumulation of phase
along some path length as in birefringence. The presence of sharp
diffraction patterns from Hubble telescope images is not
incompatible with at least simple notions of  Planck scale
physics.

 If there is some birefringence of the vacuum, that some approaches
 to quantum gravity allow, it could have had  a dramatic
  effect on any underlying
 polarization of light.
 However,  a  more realistic treatment for the effects of being constantly
buffeted by quantum gravity fluctuations shows no evidence for
random phase effects even with a underlying  ``birefringence of
the vacuum'' structure. Instead the larger the path length of
propagation the more the effects, of any underlying discreteness
of time, are weakened. There still remains a deterministic
possible alteration in the polarization that is being actively
searched for - see e.g. [7].  This makes any observation, of the
random quantum gravity influence on electromagnetic radiation from
distant sources, more difficult to envision. Some other possible
means of probing this Planck regime are recently reviewed in ref.
[9].

{\bf Acknowledgement}\\ I should like to thank Drs. R. Lieu , G.T.
van Belle and  V.A. Kosteleck{\'y} for helpful discussions.
\newpage

{\bf References}\\
\begin{enumerate}
\item R.Lieu and L.W. Hillman, Astrophysical J. Lett. 585 (2003) p.L77.\\
 also preprint  astro-ph/0301184
\item M. Born and E. Wolf, ``Principles of Optics
7th(extended) edn'', Cambridge University Press (2002).
\item L. Mandel and E. Wolf , ``Optical coherence and quantum optics''
, Cambridge University Press(1995).
\item A.S. Marathay, ``Elements of Optical coherence theory'',
John Wiley and Sons, New York. (1982).
\item G. Amelino-Camelia, Nature 398 (1999) p.216;\\
{\em ibid}, Phys. Rev. D 62 (2000) p.024015.\\ Y. Jack Ng and H.
van Dam, Found. Phys. 30 (2000) p.795.\\ also preprint
gr-qc/9906003
\item P.S. Considine, J. Opt. Soc. Am. 53 (1963) p.1351;\\
{\em ibid}, 56 (1966) p.1001.
\item V. A. Kosteleck{\'y} and M. Mewes, Phys. Rev. Lett. 87 (2001)
p.251304.\\ {\em ibid}, Phys. Rev. D 66 (2002) p.056005.
\item Y. Jack Ng, H. van Dam and W.A. Christiansen, preprint
astro-ph/0302372
\item Y. Jack Ng, preprint gr-qc/0305019
\end{enumerate}
\end{document}